\documentclass[floatfix,superscriptaddress]{revtex4-2}

\usepackage{graphicx}
\usepackage{dcolumn}
\usepackage{bm}
\usepackage{xcolor}

\begin{document}

\preprint{APS/123-QED}

\title{\textbf{Observation of Body-Centered Cubic Iron above 200 Gigapascals.} 
}%

\author{Zuzana  Kon\^{o}pkov\'{a}}
\thanks{These two authors contributed equally}
\email{Contact author: zuzana.konopkova@xfel.eu}

 \affiliation{European XFEL, Schenefeld, Germany}
 
\author{Eric  Edmund}%
\thanks{These two authors contributed equally}
\altaffiliation[Present Address: ]{%
 Universit\"at M\"unster, M\"unster, Germany
}%
 \affiliation{%
 Earth \& Planets Laboratory, Carnegie Science, 5251 Broad Branch Rd NW, Washington, DC, USA
}%

\author{Orianna B Ball}
\affiliation{
 SUPA, School of Physics and Astronomy, and Centre for Science at Extreme Conditions (CSEC), The University of Edinburgh, Peter Guthrie Tait Road, Edinburgh EH9 3FD, United Kingdom
}%

\author{Agn\`es  Dewaele}
\affiliation{%
 CEA, DAM, DIF, 91297 Arpajon, France
}%
\affiliation{%
 Universit\'e Paris-Saclay, CEA, Laboratoire Mati\`ere en Conditions Extr\^emes, 91680 Bruyères-le-Châtel, France
}%

\author{H\'el\`ene  Ginestet}
\affiliation{%
 Univ. Lille, CNRS, INRAE, Centrale Lille, UMR 8207 - UMET - Unité Mat\'eriaux et Transformations, F-59000 Lille, France
}%

\author{Rachel J Husband}
\affiliation{%
 Deutsches Elektronen-Synchrotron DESY, Notkestr. 85, 22607 Hamburg, Germany
}%

\author{Nicolas  Jaisle}
\affiliation{%
 ISTerre, Universit\'e Grenoble Alpes, CNRS, Grenoble, France
}%

\author{Cornelius  Strohm}
\affiliation{%
 Deutsches Elektronen-Synchrotron DESY, Notkestr. 85, 22607 Hamburg, Germany
}%

\author{Madden S Anae}
\affiliation{%
 Department of Geosciences, Stony Brook University, Stony Brook, NY 11794-2100
}%

\author{Daniele  Antonangeli}
\affiliation{%
Sorbonne Universit\'e, Mus\'eum National d’Histoire Naturelle, UMR CNRS 7590, Institut de Min\'eralogie, de Physique des Mat\'eriaux et de Cosmochimie (IMPMC), Paris, France.
}%

\author{Karen  Appel}
\affiliation{%
European XFEL, Schenefeld, Germany}%

\author{Marzena  Baron}
\affiliation{%
Sorbonne Universit\'e, CNRS, Laboratoire Chimie de la Mati\`ere Condens\'ee de Paris (LCMCP), 4 place Jussieu, 75005 Paris, France}%

\author{Silvia  Boccato}
\affiliation{%
Sorbonne Universit\'e, Mus\'eum National d’Histoire Naturelle, UMR CNRS 7590, Institut de Min\'eralogie, de Physique des Mat\'eriaux et de Cosmochimie (IMPMC), Paris, France.
}%

\author{Khachiwan  Buakor}
\affiliation{%
European XFEL, Schenefeld, Germany}%

\author{Julien  Chantel}
\affiliation{%
Univ. Lille, CNRS, INRAE, Centrale Lille, UMR 8207 - UMET - Unité Mat\'eriaux et Transformations, F-59000 Lille, France
}%

\author{Hyunchae  Cynn}
\affiliation{%
Lawrence Livermore National Laboratory, Physical and Life Science Directorate, Livermore, CA 94550, USA
}%

\author{Anand P Dwivedi}
\affiliation{%
European XFEL, Schenefeld, Germany}%

\author{Lars  Ehm}
\affiliation{%
Department of Geosciences, Stony Brook University, Stony Brook, NY 11794-2100
}%

\author{Konstantin  Glazyrin}
\affiliation{%
Deutsches Elektronen-Synchrotron DESY, Notkestr. 85, 22607 Hamburg, Germany
}%

\author{Heinz  Graafsma}
\affiliation{%
Deutsches Elektronen-Synchrotron DESY, Notkestr. 85, 22607 Hamburg, Germany
}%

\author{Egor  Koemets}
\affiliation{%
Department of Earth Sciences, University of Oxford, 3 South Parks Road, OX1 3AN Oxford, United Kingdom
}%
\affiliation{%
Diamond Light Source, Didcot, UK
}%

\author{Torsten  Laurus}
\affiliation{%
Deutsches Elektronen-Synchrotron DESY, Notkestr. 85, 22607 Hamburg, Germany
}%

\author{Hauke  Marquardt}
\affiliation{%
Department of Earth Sciences, University of Oxford, 3 South Parks Road, OX1 3AN Oxford, United Kingdom
}%

\author{Bernhard  Massani}
\affiliation{
 SUPA, School of Physics and Astronomy, and Centre for Science at Extreme Conditions (CSEC), The University of Edinburgh, Peter Guthrie Tait Road, Edinburgh EH9 3FD, United Kingdom
}%

\author{James D McHardy}
\affiliation{
 SUPA, School of Physics and Astronomy, and Centre for Science at Extreme Conditions (CSEC), The University of Edinburgh, Peter Guthrie Tait Road, Edinburgh EH9 3FD, United Kingdom
}%

\author{Malcolm I McMahon}
\affiliation{
 SUPA, School of Physics and Astronomy, and Centre for Science at Extreme Conditions (CSEC), The University of Edinburgh, Peter Guthrie Tait Road, Edinburgh EH9 3FD, United Kingdom
}%

\author{Vitali  Prakapenka}
\affiliation{%
Center for Advanced Radiation Sources, The University of Chicago, Chicago, Illinois 60637, USA
}%

\author{Jolanta  Sztuk-Dambietz}
 \affiliation{European XFEL, Schenefeld, Germany}

\author{Minxue Tang}
\affiliation{%
Deutsches Elektronen-Synchrotron DESY, Notkestr. 85, 22607 Hamburg, Germany
}%

\author{Tianqi  Xie}
\affiliation{%
Department of Geosciences, Stony Brook University, Stony Brook, NY 11794-2100
}%
\affiliation{%
Department of Geological Sciences, University of Saskatchewan
}%
\author{Zena  Younes}
\affiliation{
 SUPA, School of Physics and Astronomy, and Centre for Science at Extreme Conditions (CSEC), The University of Edinburgh, Peter Guthrie Tait Road, Edinburgh EH9 3FD, United Kingdom
}%

\author{Ulf  Zastrau}
 \affiliation{European XFEL, Schenefeld, Germany}

\author{Alexander F Goncharov}
\affiliation{%
Earth \& Planets Laboratory, Carnegie Science, 5251 Broad Branch Rd NW, Washington, DC, USA
}%

\author{Clemens  Prescher}
\affiliation{%
University of Freiburg, Freiburg, Germany
}%

\author{Ryan S McWilliams}
\affiliation{
 SUPA, School of Physics and Astronomy, and Centre for Science at Extreme Conditions (CSEC), The University of Edinburgh, Peter Guthrie Tait Road, Edinburgh EH9 3FD, United Kingdom
}%

\author{Guillaume  Morard}
\affiliation{%
ISTerre, Universit\'e Grenoble Alpes, CNRS, Grenoble, France
}%
\affiliation{%
Sorbonne Universit\'e, Mus\'eum National d’Histoire Naturelle, UMR CNRS 7590, Institut de Min\'eralogie, de Physique des Mat\'eriaux et de Cosmochimie (IMPMC), Paris, France.
}%

\author{S\'ebastien  Merkel}
\affiliation{%
Univ. Lille, CNRS, INRAE, Centrale Lille, UMR 8207 - UMET - Unité Mat\'eriaux et Transformations, F-59000 Lille, France
}%

\date{\today}

\begin{abstract}
The crystallographic structure of iron under extreme conditions is a key benchmark for cutting-edge experimental and numerical methods. Moreover, it plays a crucial role in understanding planetary cores, as it significantly influences the interpretation of observational data and, consequently, insights into their internal structure and dynamics. However, even the structure of pure solid iron under the Earth's core conditions remains uncertain, with the commonly expected hexagonal close-packed structure energetically competitive with various cubic lattices. In this study, iron was compressed in a diamond anvil cell to above 200 GPa, and dynamically probed near the melting point using MHz frequency X-ray pulses from the European X-ray Free Electron Laser. The emergence of an additional diffraction line at high temperatures suggests the formation of an entropically stabilized bcc structure. Rapid heating and cooling cycles captured intermediate phases, offering new insights into iron's phase transformation paths. The appearance of the bcc phase near melting at extreme pressures challenges current understanding of the iron phase diagram under Earth's core conditions. 
\end{abstract}

\maketitle


Constraining the physical properties of iron, the main constituent of Earth's core, is fundamentally important for understanding of the Earth's chemical composition\cite{Hirose2013}, thermal structure\cite{Boehler1996} and magnetic field \cite{Landeau2022}. 
Observations of elastic anisotropy in Earth's inner core have motivated numerous investigations into the phase diagram of iron at these conditions - 330 to 360 GPa and exceeding 5000 K - using a variety of experimental\cite{Brown1986,Nguyen2004,Zhang2023,Turneaure2020,Singh2023, White2020,Ping2013,Kraus2022,Tateno2010,Anzellini2013} and \textit{ab initio} techniques\cite{Stixrude1995,Stixrude2012,Belonoshko2017,Sun2023,Vocadlo2003,Zhao2024,Belonoshko2003,Belonoshko2021,Li2024}. At the heart of this question is whether or not crystalline phases other than hexagonal close-packed (hcp) iron are stable at inner core conditions.

Theoretical studies indicate that the Gibbs free energies of the various high-pressure polymorphs of iron are very similar at core conditions. The hcp structure is unambiguously the stable phase of iron at moderate temperatures above 13 GPa \cite{Stixrude1995}, however fcc and bcc structures have been proposed to become energetically competitive with the hcp structure at temperatures close to the melting point of iron at inner core pressures \cite{Stixrude2012, Belonoshko2017,Sun2023,Vocadlo2003,Zhao2024,Li2024,Belonoshko2003,Belonoshko2021}. 
The need for large simulation sizes presents a challenge, as the mechanical stability of bcc iron is highly dependent on these conditions. Self-diffusion, characterized by collective atomic motion, has been reported for both bcc iron \cite{Belonoshko2017,Ghosh2023,Li2024} and hcp iron \cite{Zhang2023} just below the melting point at high pressures. Recent \textit{ab initio} simulations augmented with machine learning have demonstrated the mechanical stability of the bcc structure of pure iron under Earth's core conditions. However, its thermodynamic stability compared to an hcp structure appears to be enhanced only in the presence of impurities \cite{Li2024}.

Additional experimental investigations into the structure of solid iron at Earth's core conditions are required, particularly at temperatures close to melting, but measuring the structure of iron at such conditions is experimentally challenging. 
Dynamic compression of iron provides one method to investigate iron’s phase stability above 200 GPa. Measurements of crystal structure have been carried out in single shock experiments\cite{Turneaure2020, Singh2023, White2020, Balugani2024}, and in more complex compression experiments\cite{Kraus2022}. While these studies observe only an hcp structure up to melting, reaction kinetics and the shifting of phase stabilities due to high strain rates and short experimental timescales can potentially lead to discrepancies between phase transitions encountered during dynamic compression and static compression experiments\cite{Smith2013,Gorman2018}.
 
Static compression experiments offer an alternative avenue to investigate the structure of iron at core conditions\cite{Tateno2010, Kuwayama2008, Anzellini2013, Konopkova2021, Sinmyo2019}. However, despite being able to access longer timescales and hence better approach thermodynamic equilibrium conditions, static compression experiments suffer from challenges due to the large temperature gradients present in the samples, and the difficulties of avoiding chemical contamination of the compressed iron sample \cite{Aprilis2019,Tateno2010,Morard2018a}.

These challenges necessitate the use of new experimental tools and capabilities with which to probe phase transitions at extreme conditions. Experimental timescales need to be sufficiently short to suppress unwanted chemical reactions, while being sufficiently long to minimize the effects of transition kinetics; highly monochromatic X-rays which probe the sample with high time resolution are needed to better resolve structural transition details, including difficult to detect structural intermediary states. 
Such experiments have been carried out at the High Energy Density (HED) beamline of the European X-ray Free Electron Laser (EuXFEL) using the diamond anvil cell (DAC) platform\cite{Liermann2021,Zastrau2021,Ball2023}.

\begin{figure}%
\centering
\includegraphics[width=0.7\textwidth]{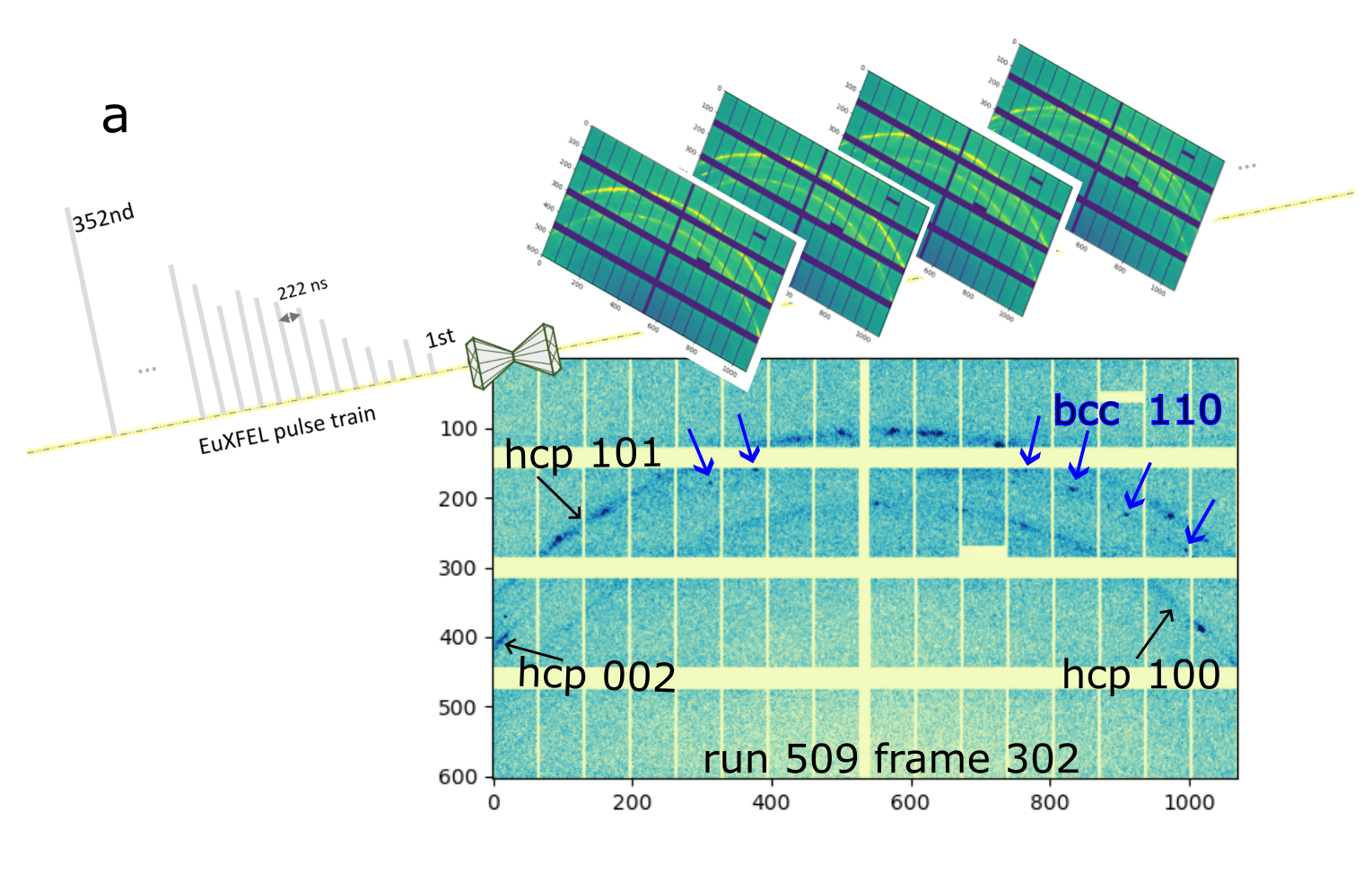}
\includegraphics[width=0.8\textwidth]{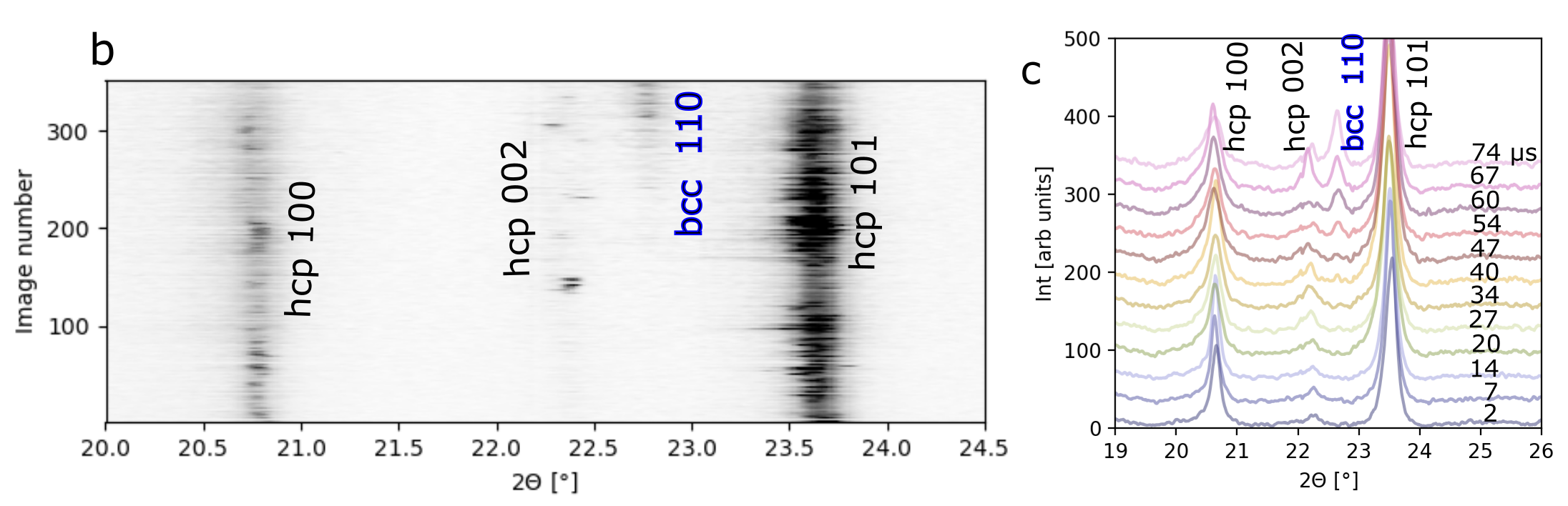}
\includegraphics[width=0.8\textwidth]{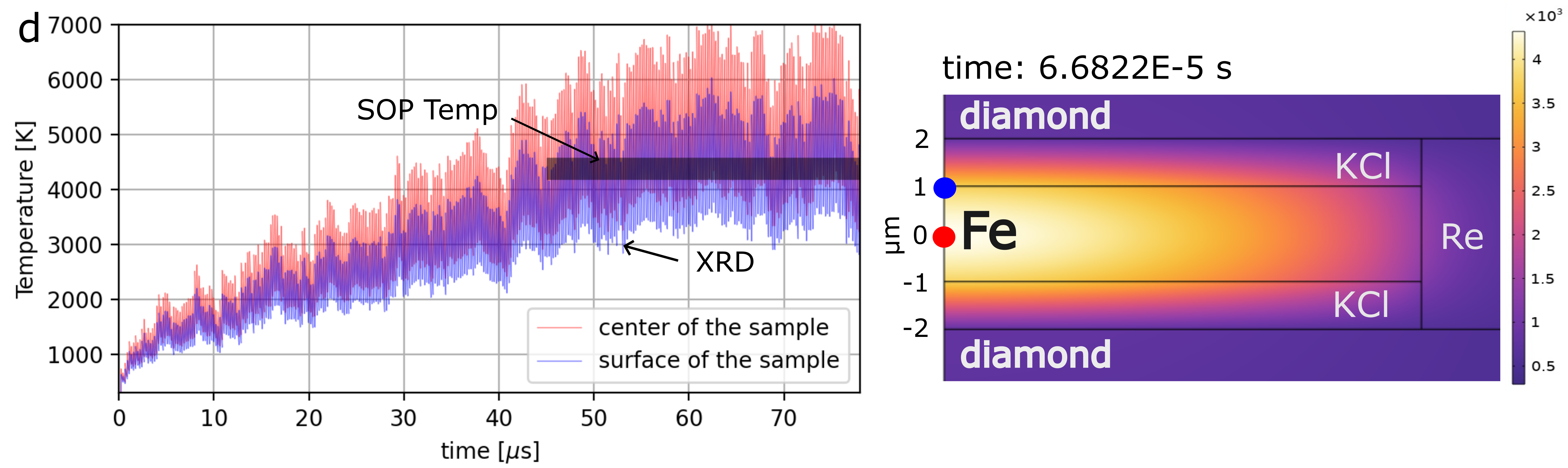}
\caption{\textbf{Data showing the first pulse train (Sample CC273, run 509) that transforms part of the sample into the cubic structure.} \textbf{a}, Sketch of the experiment. A raw diffraction image from pulse 302 displays several spots corresponding to the 110 bcc-Fe reflection (blue arrows). \textbf{b}, Stacked integrated diffraction patterns from all 352 pulses show a new reflection towards the end of the train (time from bottom-up), coinciding with a decrease in intensity of all hcp-Fe peaks. \textbf{c}, 2$\theta$ vs. intensity plots of the integrated patterns, averaged over 10 pulses in sequences of every 30 pulse (i.e. average of 1-10, 30-40, ..., 330-340; from bottom to top) showing the emerging 110 bcc-Fe reflection. The patterns are offset vertically for clarity. \textbf{d}, An FEA model illustrates the spatial and temporal evolution of temperature (from the center and surface of the sample, marked by red and blue points, respectively) after each pulse. X-ray diffraction (XRD) is measured at the minimum temperature of each oscillation. The dark line indicates the average surface temperature of the sample, measured by streaked optical pyrometry (SOP).}
\label{Fig 509}
\end{figure}

Pure iron foils, embedded within a KCl pressure-transmitting medium, were compressed in two DACs to pressures of 218 GPa and 208 GPa, respectively, and exposed to a train of short, $\sim$20 fs X-ray laser pulses. Each pulse train consists of 352 X-ray pulses at a 4.5 MHz repetition rate (i.e. 222 ns separation between pulses, 78 $\mu$s train duration). Each time the sample is exposed to an X-ray pulse, the pulse is partially absorbed, causing the sample to heat up over time as the rate of heating by X-ray absorption exceeds the rate of heat diffusion from the sample. The X-ray pulse intensity has been chosen to sequentially increase over the duration of the pulse train to gradually heat the sample (Fig.~\ref{Fig 509}A). Due to the nature of X-ray heating by short X-ray pulses, each diffraction pattern in a pulse train reflects the sample's structure on cooling from the previous pulse, as heating occurs after the immediate scattering.

The initial low intensity X-ray pulse trains produced diffraction patterns showing only the 100, 002 and 101 peaks from the hcp structure of iron (Supplemental Material, Fig.~\ref{pre-BCC}). As the X-ray pulse intensity increases throughout the pulse train, the rising temperature is indicated by diffraction peaks shifting to lower diffraction angles due to thermal expansion of the crystal lattice. 
When pulse train intensity was increased, an additional peak appeared between the positions of the 002 and 101 reflections of hcp iron during the last third of the pulse train (Fig.~\ref{Fig 509}B,C, Fig~\ref{Fig 553}, movie 1 in Supplemental Material). This peak is absent in the cold starting material and is not observed in the diffraction data after temperature quench (Supplemental Material, Fig.~\ref{before_after_bcc}). Aside from the new phase, no peaks can be attributed to materials other than the starting components in the DAC sample chamber - hcp-Re, bcc-KCl and hcp-Fe. Potential contaminants, such as carbides (e.g. Fe$_3$C\cite{Aprilis2019,Tateno2010,Morard2018a}), can be excluded due to their structural dissimilarity to the observed diffraction patterns and the fact that the new reflection was not observed after quenching from high temperatures which is not expected for structures resulting from chemical reactions. The total combined duration of sample heating by the pulse trains prior to the phase transition remained well under 1 ms.

\begin{figure}%
\centering
\includegraphics[width=0.75\textwidth]{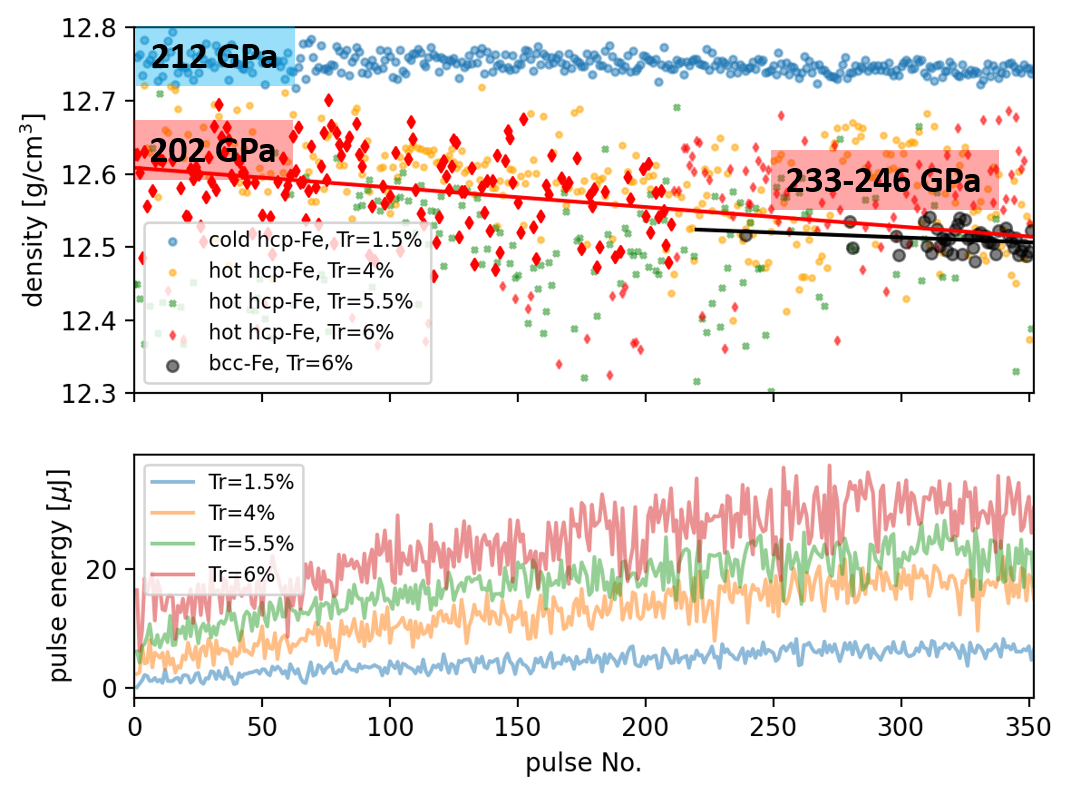}
\caption{\textbf{Calculated densities of hcp and bcc phases.} Data are from runs 500 (1.5\% transmission), 505 (4\%), 508 (5.5\%) and 509 (6\%) (Sample CC273, top panel) with increasing pulse energies (bottom panel), demonstrating that the volume change between the hcp and bcc structures is small but positive. In red is the first run in the series with the bcc structure. Red and black lines are linear regressions to hcp and bcc densities, respectively. 'Tr' indicates X-ray transmission.
}
  
\label{Vols}
\end{figure}

The individual Bragg reflections of the new phase display both large crystallite spots and a powder component. The number and intensity of the observed bcc spots increase while the average intensity of all three hcp reflections decreases, indicating that the new phase results from a bulk transformation of the iron foil. 
Given that only a single new peak is detected over the whole observed 2$\theta$ range, the most plausible structure is one with high symmetry. Indexing this peak as the 111 fcc reflection results in a volume that is 7\% lower than the ambient temperature hcp structure, which is too large a difference to be plausible\cite{Mikhaylushkin2007}. Additionally, the second fcc reflection, 200, which would still fall within the detectable diffraction angular range, is never observed. Indexing the new peak to the 110 reflection of the bcc structure is the only solution that provides a physically realistic volume (1.3\% higher than 300 K hcp) and matches the observed number of peaks, with the second bcc reflection, 200, falling outside the range of the detector. The volumes of coexisting hcp and bcc phases are roughly the same at the transformation temperature (Fig.~\ref{Vols}), indicating that the densities are similar. The data is consistent with a positive volume change between hcp and bcc phases, which implies a flat or positive Clapeyron slope (dT/dP = $\Delta$V/$\Delta$S) for the hcp-bcc transition. A positive Clapeyron slope indicates that the temperatures required to stabilize this phase would increase with increasing pressure.

The average temperature during the X-ray pulse train, where thermal emission become detectable and coincides with observation of the bcc phase, is estimated to be 4400(500) K based on optical pyrometry measurements\cite{Ball2023} (Fig.~\ref{SOP_r553}). Finite element analysis (FEA) models were aligned with observed temperatures to match the spatial-temporal average temperature from the surface of the sample during the final third of the pulse train. FEA modeling\cite{MezaGalvez2020} shows that the hottest region of the sample is at its center while it undergoes repetitive heating and cooling cycles, with temperature oscillations increasing as the absolute temperature rises (Fig.~\ref{Fig 509}D). Notably, the X-ray diffraction (XRD) snapshot is always taken at the minimum temperature of the cycle, i.e. after cooling from peak temperatures. Temperature can decrease by as much as 2000 K between the peak temperature and the X-ray measurement at X-ray intensities where the new peak is observed (Fig.~\ref{Fig 509}D), implying that part of the sample may have cooled from a liquid state\cite{Anzellini2013}.

The experiment was conducted twice with similar observations from starting pressures of 208 GPa (HIBEF38, runs 517-562 (Fig.~\ref{Fig 553})) and 218 GPa (CC273, runs 498-516), although it is noted that pressure varied during the experiments due to the increase in thermal pressure during heating and the decrease in pressure due to annealing during initial runs (Fig.~\ref{pre-BCC}). 
Other samples tested at lower pressures (130-180 GPa starting pressure, see Methods) reveal a more complex behavior. Initially, the diffraction patterns are dominated by strong streaks connecting the 100 and 101 hcp-Fe lines (Fig.~\ref{SF}B, SI movie 2), typically attributed to stacking faults in the hcp-fcc structure, as also observed in standard laser heating experiments\cite{Konopkova2021}. We also observed incipient nucleation of the bcc structure, indicated by the presence of diffuse speckles and occasional isolated bcc spots (Fig.~\ref{SF}A,C).
This implies that the hcp, bcc, and fcc phases are in competition, with the bcc phase being transient and not developing into a fully crystalline phase that could be unambiguously detected in the diffraction data below 180 GPa.
   
\begin{figure}
    \centering
    \includegraphics[width=0.9\linewidth]{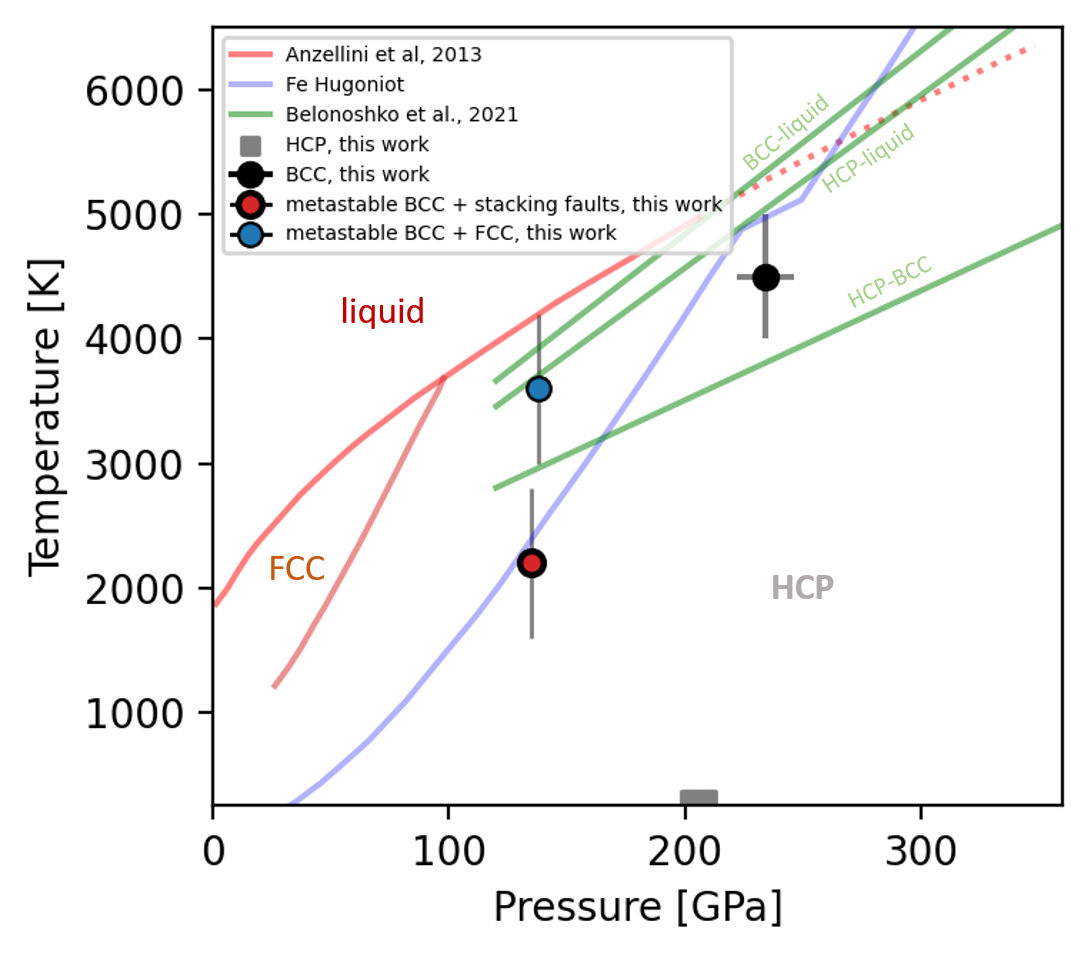}
    \caption{\textbf{Phase diagram of iron.} Our experimental data are depicted as circles, with colors indicating the observed structures in the diffraction data: black for bcc, blue for fcc, red for stacking faults in hcp/fcc, and grey for pure hcp (always present). Bright red lines represent the melting curve, the fcc stability region based on the laser heating DAC study by Anzellini et al.\cite{Anzellini2013}. Green lines indicate the hcp-bcc, hcp-liquid, and bcc-liquid boundaries computed from a molecular dynamics study by Belonoshko et al.\cite{Belonoshko2021}. The blue line is the Fe Hugoniot from Ref. \cite{Brown1986}. The red and blue points are possible phases stabilized by solidification kinetics, as computed by Sadigh et al.\cite{Sadigh2021}. }
    \label{fig:phase_diag}
\end{figure}

Our observations are summarized in Fig. \ref{fig:phase_diag}. 
At pressures above 200 GPa, the bcc structure is clearly stabilized at high temperatures below the melting point upon cooling from higher temperature, hence possibly forming during the solidification pathway from the melt. At pressures below 180 GPa, the data suggest competition between metastable bcc crystallite formation and thermally disordered closed-packed phases at high temperatures, consistent with previous reports\cite{Mikhaylushkin2007,Konopkova2021}. 

\begin{figure}%
\centering
\includegraphics[width=0.7\textwidth]{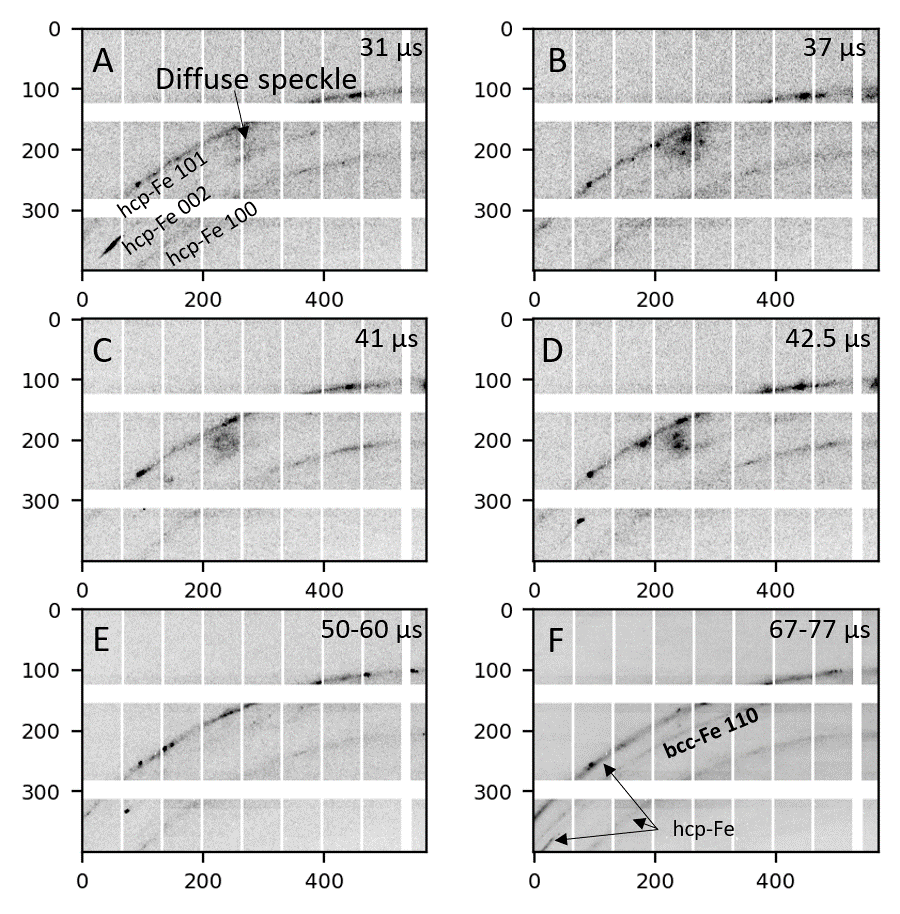}
\caption{\textbf{Evolution of diffuse speckles to powder-like ring of 110 bcc-Fe reflection with time.} Sample CC273, Run 509. \textbf{a}, Diffuse speckles of bcc-Fe appear $\approx$30 $\mu$s into the run, between the 002 and 101 diffraction lines of hcp-Fe. \textbf{b}-\textbf{e}, Bcc speckles evolve into more localized, sharper spots and shift in azimuth and towards a slightly smaller $2\theta$ between $\approx$30 and $\approx$60 $\mu$s. \textbf{f}, Later in the run, bcc-Fe is fully crystallized and forms a continuous powder diffraction ring.
}
\label{speckle}
\end{figure}

The time-resolved nature of the experiment allows us to observe the formation of the first bcc grains as the temperature gradually increases (Fig. \ref{speckle}). Initially, isolated diffuse speckles appear about 31 $\mu$s into the train at the 2$\theta$ angle between the 002 and 101 hcp peaks (Fig. \ref{speckle}A-D). The temperature at the probing moment is about 3000 K.  
As the pulse train progresses, these broad speckles evolve into more localized sharper spots and shift to a slightly smaller 2$\theta$ angle (3500-4500 K, 37-43 $\mu$s). Up to this point, the temperature oscillations should remain below the melting point of iron at this pressure, approximately 5200 K. At the end of the train (55-77 $\mu$s, 5000 K), very sharp, bright Bragg spots form with a random distribution along the azimuthal angle (Fig. \ref{speckle}F). During this time, the maximum temperatures most likely exceed the melting line. A plausible mechanism for this phase transformation is the formation of a bcc nanotwinned structure, followed by de-twinning, as recently simulated in molecular dynamics studies of Mg\cite{Zhou2023}. In this mechanism, the bcc growth is facilitated by formation of nano-twins by atomic shear-shuffle that efficiently accommodate the shear caused by the formation of a small bcc seed within the hcp matrix. 
The nano-twin region exhibits a lamellar structure oriented along the c-axis of the hcp lattice, which could account for the texture of the 002 hcp reflection observed at the same azimuthal angle as the diffuse spots. The small size of the bcc nanotwins and their transitory nature align with the observed diffuse, single-crystal-like speckles that appear briefly between consecutive X-ray pulses. Subsequent de-twinning (aided by possible melting) leads to the formation of polycrystalline bcc Fe and the further growth of a relatively large number of grains, as evidenced by the final multiple randomly distributed spots in our diffraction patterns.

An important question arises as to why this long-sought-after phase has not been observed in previous experimental studies on iron.  This may be related to the thermodynamic pathways investigated in the present study. The MHz heating and cooling cycles used to generate extreme temperatures here are numerically simulated to result in cooling rates of $\sim$10 K/ns when temperatures exceed 4000 K. Molecular dynamics calculations of rapid solidification of liquid copper at comparable cooling rates (100 K/ns) have shown that metastable high temperature phases can be formed at the interface between the liquid and the nucleating solid phase as this configuration reduces the free energy of the interface\cite{Sadigh2021}.
The polymorphic solidification observed in their study matches our findings: below a certain pressure (in our study below 180 GPa), small body-centered cubic nuclei (bcc) and close-packed nuclei form, but the bcc nuclei quickly transform into close-packed phases (fcc, hcp), while above this pressure (200 GPa in our study), nucleation starts in the bcc phase and grows, suggesting that the solidification process stabilizes the metastable bcc phase. Such kinetically stabilized phase regions for iron implied by our experimental observations are depicted in Fig. \ref{fig:phase_diag}: the red circle marks conditions where unstable bcc nucleation competes with the formation of hcp/fcc stacking faults at lower temperatures, while fcc growth is favored at higher temperatures (blue circle). At higher pressures (black circle), 
the final powder texture of the bcc phase appearing in many consecutive XRD images above 230 GPa suggests that a narrow thermodynamic stability field of the bcc phase is reached. 

Compared to standard laser heated (LH) DAC experiments, our methodology offers significant advantages in detecting new phases. The high rate of diffraction data (352 XRD patterns in 78 $\mu$s) and high brilliance monochromatic source coupled to novel volumetric heating enable unambiguous detection of structural intermediary states and conditions directly adjacent to melting. The rapid dynamics of nucleation could be challenging to observe in synchrotron XRD experiments.
The volumetric absorption of the X-ray energy helps transform a larger sample volume in the center of the foil, even as the surface remains close to the cool diamond interface, particularly at extreme pressures. 
In addition, X-ray heating experiments always probe high temperature states upon cooling, and therefore access different thermodynamic pathways from most conventional LH DAC and laser compression experiments which study phase transitions upon heating \cite{Tateno2010,Anzellini2013,Turneaure2020,Singh2023,White2020}. 
Our study probes the behaviour of iron at timescales intermediate to laser compression, $\sim$1-10 ns, and LH DAC experiments - milliseconds to minutes; MHz X-ray heating experiments are most similar in timescale to plate impact experiments ($\sim$100-1000 ns), where there is no consensus on the phase diagram between 200-260 GPa\cite{Brown1986,Nguyen2004,Zhang2023}. 

A key distinction between our experiments and previous studies lies in the nature of heating. In our case, heating is induced by intense, short-pulse hard X-ray irradiation at power densities of $10^{15}-10^{16}$ W/cm$^2$. This regime enables direct core-electron excitation, leading to extremely high electron temperatures within the first few picoseconds. Rapid rise in electron temperature can drive nonthermal phase transitions by altering ionic potentials before significant lattice heating occurs.
For instance, in iron at ambient pressure, theoretical studies predict that a solid-solid phase transition from the bcc to the fcc structure can be induced under high electron temperatures, even when the lattice remains near room temperature\cite{Azadi2024}. This transformation is attributed to phonon softening facilitated by ultrafast demagnetization. However, a crucial question remains: can a phase transition initiated under such extreme electronic conditions persist over longer timescales, particularly after electron-ion energy equilibration? This is especially relevant in cases where the Gibbs free energies of competing high-pressure phases are similar, making their stability highly sensitive to transient electronic effects\cite{Ghosh2023}.

Our results place the formation of the bcc structure in the range of 233 - 246 GPa and 4000-5000 K, based on the measured hcp-Fe volumes and the thermal equation of state of Fe\cite{Dewaele2006} (Fig. \ref{fig:phase_diag}, Table \ref{tab:EoS}). If the cubic structure is indeed the stable phase, it has significant implications for the structure of Earth's solid inner core.
The presence of an intermediate structure between hcp Fe and liquid Fe supports a higher melting temperature estimate at inner core conditions. This is because the entropy change across the hcp-bcc and bcc-liquid transitions must be equal to or larger than the entropy change of the lower pressure hcp-liquid transition\cite{Bassett1990,Komabayashi2010}. \textit{Ab initio} calculations propose that magnetic and/or configurational entropy are enhanced for the bcc structure at inner core conditions, aiding in its stabilization at these conditions\cite{Belonoshko2017,Pourovskii2013,Ghosh2023}.

Further, the composition of Earth’s inner core is governed by the partitioning of light elements into solid iron during crystallization. The solubility of light elements in iron will change under these conditions due to symmetry differences between hcp and bcc lattices. Most studies on light element partitioning in solid and liquid iron using \textit{ab initio} calculations assume the hcp structure as stable at inner core conditions\cite{Alfe2000,Li2019}, and experimental studies often do not reach the stability field of bcc iron observed here\cite{Ozawa2016,Miozzi2020,Tagawa2022}, indicating that partitioning results could differ significantly from expectations at Earth’s inner core conditions. Previous experimental and theoretical studies of the elasticity of iron alloys, used to infer inner core composition, are predominantly focused on the properties of hcp alloys\cite{Edmund2019,Huang2022a,Martorell2016,He2022}, and so the stabilization of a different phase at these conditions calls into question existing compositional models for the Earth's inner and outer core.

One of the biggest open questions is the large-scale anisotropic features in the inner core of Earth, where compressional waves travel faster along the rotation axis, with distinct hemispherical differences, and with a depth-dependent increase in anisotropy, reaching up to 8\% at the core's center\cite{Lythgoe2014,Frost2021,CostadeLima2022}. While the outermost layer of the inner core appears isotropic at large scales, local-scale anisotropy has been detected\cite{Tkalcic2024}, with suggestions of coexisting bcc and hcp iron phases near the inner core boundary\cite{Zhang2023}. Numerous models attempt to explain these anisotropic structures, incorporating both hcp and bcc phases of iron\cite{Mattesini2013,Frost2021}. However, while bcc Fe exhibit high anisotropy at the crystal level\cite{Ghosh2023}, aligning this with the coherent, large-scale anisotropic structures seen seismologically remains challenging to reconcile with plastic deformation induced by large-scale flow in bcc-Fe\cite{Lincot2015}. Our  findings indicate that bcc-Fe can form upon cooling from molten Fe at the inner-core boundary, or may have a narrow stability field near melting at core pressures. This scenario aligns with observed small-scale anisotropy and the absence of large-scale anisotropic features in the outer shell of the inner core. At greater depths, reproducing the global anisotropy of the inner core would require substantial  alignment of bcc crystals\cite{Mattesini2013}, which is inconsistent with large scale flow\cite{Lincot2015}. Over geological timescales, Fe might revert to the hcp structure in deeper core regions, enabling the formation of anisotropic structures through flow, or the inner core might retain its crystallization structure upon freezing. This will have to be resolved with further studies.

\begin{acknowledgments}
The authors are indebted to the HIBEF user consortium for the provision of instrumentation and staff that enabled this experiment. We acknowledge European XFEL in Schenefeld, Germany, for provision of X‐ray free‐electron laser beamtime at Scientific Instrument HED (High Energy Density Science) and would like to thank the staff for their assistance. We acknowledge DESY (Hamburg, Germany), a member of the Helmholtz Association HGF, for the provision of experimental facilities. Parts of this research were carried out at PETRA III (beamline P02.2). The authors also wish to thank S. Chariton for assistance during beamtime. The authors wish to thank S. Vitale for technical assistance with Focused Ion Beam milling at Earth \& Planets Laboratory. We thank H.P. Liermann, T. Michelat, P. Schmidt and J. Moore for their support.
\end{acknowledgments}

\section*{Author contributions}
Conceptualization	Z.K., E.E., H.C., H.M., C.P., R.S.M., G.M., S.M.
Methodology	Z.K., N.J., C.S., K.A., H.G., T.L., H.M., M.I.M., J.S., C.P., R.S.M., G.M., S.M.
Software	Z.K., O.B.B., C.P.
Validation	Z.K., H.G., N.J., C.S., C.P., R.S.M., G.M., S.M.
Formal analysis	Z.K., E.E., A.D., H.G., N.J., E.K., V.P., R.S.M., S.M.
Investigation	Z.K., E.E., A.D., R.J.H., N.J., C.S., A.S.M., K.A., M.B., S.B., K.B., J.C., A.P.D., L.E., K.G., E.K., B.M., J.D.M., M.I.M., V.P., M.T, T.X., Z.Y., A.F.G., R.S.M., S.M.
Resources	Z.K., A.D., C.S., D.A., L.E., E.K., H.M., M.I.M., A.F.G., R.S.M., S.M.
Data Curation	Z.K., E.E., N.J., A.F.G., C.P., G.M., S.M.
Writing - Original Draft	Z.K., E.E., G.M.
Writing - Review \& Editing	Z.K., E.E., A.D., N.J., C.S., D.A., S.B., H.C., H.M., B.M., J.D.M., M.I.M., Z.Y., A.F.G., C.P., R.S.M., G.M., S.M.
Visualization	Z.K., E.E., A.D.
Supervision	Z.K., E.E., C.S., H.M., A.F.G., R.S.M., G.M., S.M.
Project administration	Z.K., H.M., U.Z., A.F.G., C.P., G.M., S.M.
Funding acquisition	D.A., K.A., H.M., U.Z., A.F.G., R.S.M., S.M.

\section*{Data}

The original European XFEL data is located at \url{https://in.xfel.eu/metadata/doi/10.22003/XFEL.EU-DATA-003063-00}, and will be publicly available in 2025 after the embargo period of 3 years. Source data for figures and intermediate data processing results are provided at \url{https://nextcloud.univ-lille.fr/index.php/s/pWKnP8irdNYbDcb} and will be made available on Zenodo after acceptance of the paper.

\section*{Funding}


D.A. and S.B. have received funding from the European Research Council (ERC) under the European Union’s Horizon 2020 research and innovation Programme (Grant agreement 724690). 
H.M. and E.K. were supported by European Union’s Horizon 2020 research and innovation Programme (ERC grant 864877) and UKRI STFC grant ST/V000527/1.
R.S.M., B.M., and Z.Y. were supported by the ERC under the European Union’s Horizon 2020 research and innovation Programme (Grant agreement 101002868) and RCUK- EPSRC Grant Nos. EP/P024513/1.
S.M., J.C, and H.G. are funded by the European Union (ERC, HotCores, Grant No 101054994). 
Views and opinions expressed are however those of the author(s) only and do not necessarily reflect those of the European Union or the European Research Council. Neither the European Union nor the granting authority can be held responsible for them.
K.A. received support from DFG grant AP262/2-2 and K. B. received funds from DFG project AP262/2-2.
This work was supported by Grant No. EP/S022155/1 (M.I.M., M.J.D.) from the UK Engineering and Physical Sciences Research Council. J.D.M. is grateful to AWE for the award of CASE Studentship P030463429.
O.B.B. acknowledges the support of the Scottish Doctoral Training Program in Condensed Matter Physics (CM-CDT).
G.M. and N.J. were supported by ANR grant MinDIXI (ANR-22-CE49-0006) and Labex OSUG@2020 (Investissements d’avenir—ANR10 LABX56) and IDEX Université Grenoble Alpes.
L.E., A.S.M, and T.X. were supported by NASA–Solar-System-Workings program under grant number 80NSSC17K0765.
A.F.G. and E.E. were supported by US National Science Foundation Grant EAR‐ 2049127, CHE 2302437, and Carnegie Science.
Part of this work was performed under the auspices of the U.S. Department of Energy by Lawrence Livermore National Laboratory under Contract DE-AC52-07NA27344 (H.C.).	
V.P. was supported by the National Science Foundation-Earth Sciences (EAR-658 1634415).

\newpage
\section*{Supplemental Material}

The experiments were part of the \#3063 community proposal carried out at the High Energy Density (HED) beamline of the European X-ray Free Electron Laser (EuXFEL) using the diamond anvil cell (DAC) platform\cite{Liermann2021,Zastrau2021,Ball2023}. 
17 diamond anvil cells were prepared with starting pressures ranging from 10 to 218~GPa, with rhenium gaskets, and either KCl or SiO$_2$ as pressure medium. Samples were made from iron foils of 5~$\mu$m thickness (Sigma-Aldrich, purity 99.85\%, part \#GF19019000) cut into disks of 10, 20, or 40~$\mu$m diameter using a femtosecond laser at CEA, Bruyeres-le-Chatel, France, and distributed among team members. Among those, 5 samples were studied at pressures above 120~GPa:
CDMVA9 and CDMX10, both prepared at CEA, Bruyeres-le-Chatel, France, with starting pressures of 134 and 122 GPa, respectively, BetsaA at Univ. Lille, France, with a starting pressure of 152 GPa, 
HIBEF38 at EuXFEL in Hamburg, Germany, with a starting pressure of 208~GPa, and CC273 at the Earth \& Planets Laboratory, Carnegie Institution for Science, Washington, DC, with a starting pressure of 218~GPa. All cells were checked for X-ray diffraction and a starting pressure estimate at the P02.2 beamline of PETRA III, Hamburg, Germany, using a monochromatic X-ray wavelength of 0.2909~$\AA$ a Perkin-Elmer XRD 1621 flat panel detector with $2048 \times 2048$ pixels of $0.2 \times 0.2$~mm$^2$ placed at a distance of 416~mm from the sample. Pressures were estimated using the equation of state of Fe \cite{Dewaele2006}.

HIBEF38 used two identical beveled diamond anvils with 40 $\mu$m culets, TYPE Ia, (100)-oriented, standard design. A 10-15 $\mu$m diameter pressure hole was drilled in the center of an indented rhenium gasket using a spark-erosion (EDM) machine. A disc of pure iron foil was placed between two discs of KCl in the gasket hole with the help of a micromanipulator. The assembly was then pressurized to 208 GPa, judged by XRD during pre-analysis beamtime.

CC273 used toroidal diamond anvils shaped using a focused ion beam. The toroidal anvils had culet diameters of 40 $\mu$m, and the sample chamber was composed of a Re foil, initially 250 $\mu$m thick, indented to a central thickness of 15 $\mu$m. A hole of about 25 $\mu$m diameter was then drilled using a sub-nanosecond laser. This cell was loaded with KCl and Fe plates, and then left in a vacuum oven at 110 $^{\circ}$C and a pressure of less than 1 mbar for 24 hours. After this time, the cell was sealed and pressure was raised to 50 GPa, measured using Raman spectroscopy of the T$_{2g}$ diamond phonon\cite{Akahama2006}, for transportation to Petra III. At pressures of a few GPa, Raman measurements of the sample chamber at a wavelength range of 2800-3500 cm$^{-1}$ did not show evidence of O-H stretching modes, indicating that moisture contamination of the sample chamber was below detection limit of the technique. The pressure was increased to 218 GPa at P02.2. 

BetsaA used two identical conical beveled diamond anvils with 100 $\mu$m culets, TYPE Ia, (100)-oriented, with a 70$^\circ$ aperture. A 50 $\mu$m diameter pressure hole was drilled in the center of an indented rhenium gasket using a picosecond laser drill. A 20 $\mu$ diameter Fe disc was placed between KCl in the gasket hole, KCl used as pressure medium was stored at 110$^{\circ}$C prior to loading. The assembly was then pressurized to 152 GPa.

CDMVA9 and CDMX10 used beveled diamond anvils with 70 and 100~$\mu$m culets, respectively. Both cells were prepared with Re gaskets pre-indented to a thickness of 17 $\mu$m (CDMVa9) and 19 $\mu$m (CDMX10), from a 250 $\mu$m thick Re foil. Fe samples were prepared by thinning Fe to a thickness of 2-3 $\mu$m. KCl was loaded as disks, prepared with a femtosecond laser drill. The starting KCl powder was stored at 350$^{\circ}$C prior to preparation of the KCl disks.  After loading, the pressure was increased to 134 GPa in CDMVa9 and 122 GPa in CDMX10 according to the equation of state of iron\cite{Dewaele2006}. CDMVa9 and CDMX10 were measured at the EuXFEL at a 4.5 MHz X-ray pulse repetition rate. The pressure was then raised on CDMVa9 to 172 GPa and again measured at a 2.2 MHz X-ray pulse repetition rate. 2.2 MHz repetition rate experiments demonstrated qualitatively similar behaviour to lower pressure runs, but the reduced time resolution hindered quantitative assignment of transition temperatures. 

X-ray pulse trains at 18 keV were delivered into the sample chamber focused to a 4-7 $\mu$m FWHM spot size, partially absorbed along the pathway through the DAC. X-ray diffraction patterns were collected for each pulse using an adaptive gain integrated panel detector (AGIPD). Each pulse train comprised 352 individual pulses spaced 222 ns apart, with each X-ray pulse having a pulse-width of less than 30 fs\cite{Zastrau2021}. The 4.5 MHz repetition rate of the X-ray pulses led to gradual heating of the sample over the 78 µs duration of the pulse train. The evolution of the sample's structure was monitored through recorded X-ray diffraction patterns. The femtosecond pulse-width of the X-ray pulses generated at EuXFEL enabled atomic-scale snapshots of the materials within the DAC, with collection times shorter than the typical timescale of lattice vibrations (approximately 1 ps). This experimental setup was complemented by streaked optical pyrometry (SOP), which captured time-resolved thermal emission to monitor the temperature evolution of the sample over time\cite{Ball2023}.

After aligning the DAC sample in the X-ray beam and for thermal emission measurement, a single highly attenuated X-ray pulse train was used to probe the sample without significantly increasing its temperature. These pulse trains consisted of X-ray pulses with increasing intensity towards the end of each train (Fig. \ref{pre-BCC}B, bottom panel), allowing initial probing of the sample in its cold state or with slight annealing. For each sample, multiple runs were acquired as the X-ray beam transmission was increased incrementally (typically in steps of 0.5 or 1 percent), and the sample was probed again at the same location. Figure \ref{pre-BCC}B illustrates the evolution of the sample, observed through the 101 hcp-Fe peak position across multiple pulse trains with increasing intensity. After the first couple of trains, the 2$\theta$ position of the Fe peaks shifted to lower angles, indicating annealing of the sample due to elevated temperatures during preceding pulse trains. The first train (run 500 (CC273) and run 538 (HIBEF38)) in Fig. \ref{pre-BCC} shows slight temperature increase towards the end, while more significant heating is evident in the next trains (run 505 (CC273) and run 550 (HIBEF38)), marked by the first appearance of the hcp-Fe 002 peak. In subsequent higher-power trains, additional single-crystal-like 002 Bragg peaks appeared, yet the position of the 101 hcp-peak ceased to shift towards lower 2$\theta$ values with increasing X-ray pulse energies.

\renewcommand{\thefigure}{S\arabic{figure}}
\setcounter{figure}{0}
\begin{figure}[h]%
\centering
\includegraphics[width=0.99\textwidth]{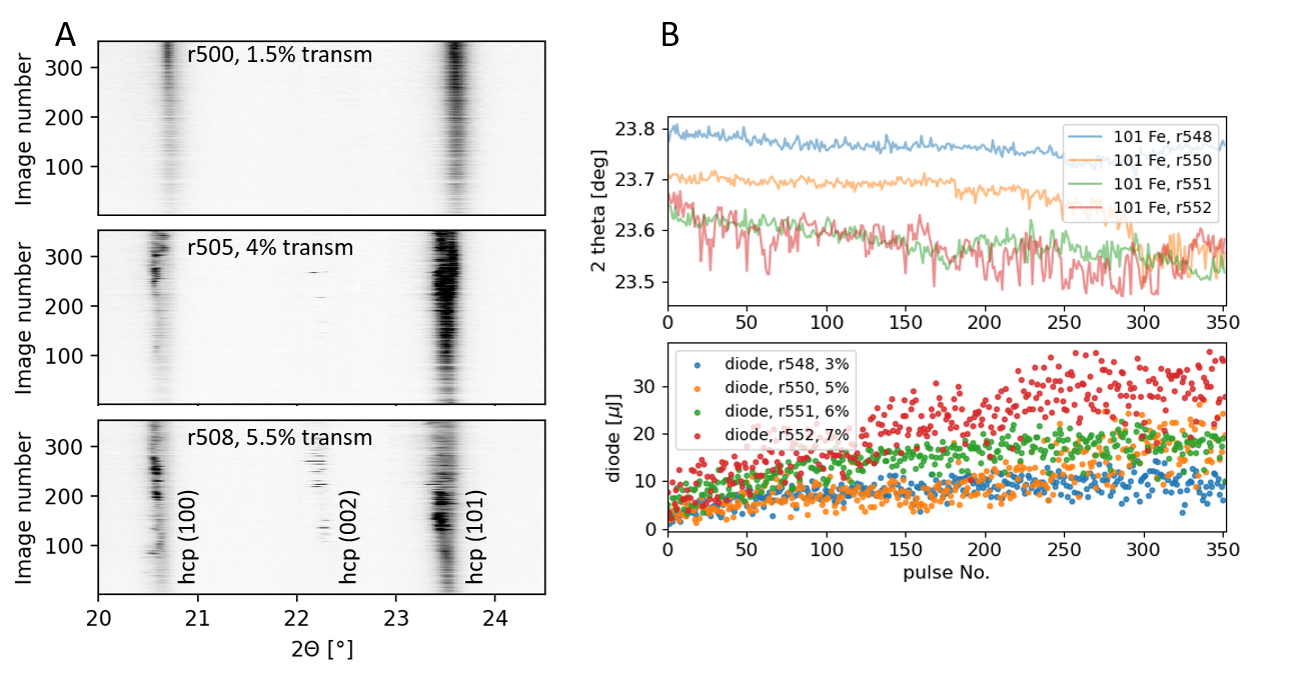}

\caption{\textbf{Data from runs preceding to bcc-Fe appearance.} \textbf{a}, Stacked integrated diffraction patterns from the CC273 DAC, with successively increasing X-ray transmission (i.e. increasing pulse energies) from the top panel to the bottom. Re-crystallization begins at above 4\% transmission, indicated by the more frequent occurrence of the 002 hcp-Fe reflection. \textbf{b}, HIBEF38 DAC: The 2$\theta$ position of 101 hcp-Fe peak during successive trains with increasing X-ray energies shows annealing of the starting pressure indicated by lower 2$\theta$ value of the first diffraction of the train, stabilizing at 23.65$^{\circ}$ (green and red data).}\label{pre-BCC}
\end{figure}

\begin{figure}%
\centering
\includegraphics[width=0.9\textwidth]{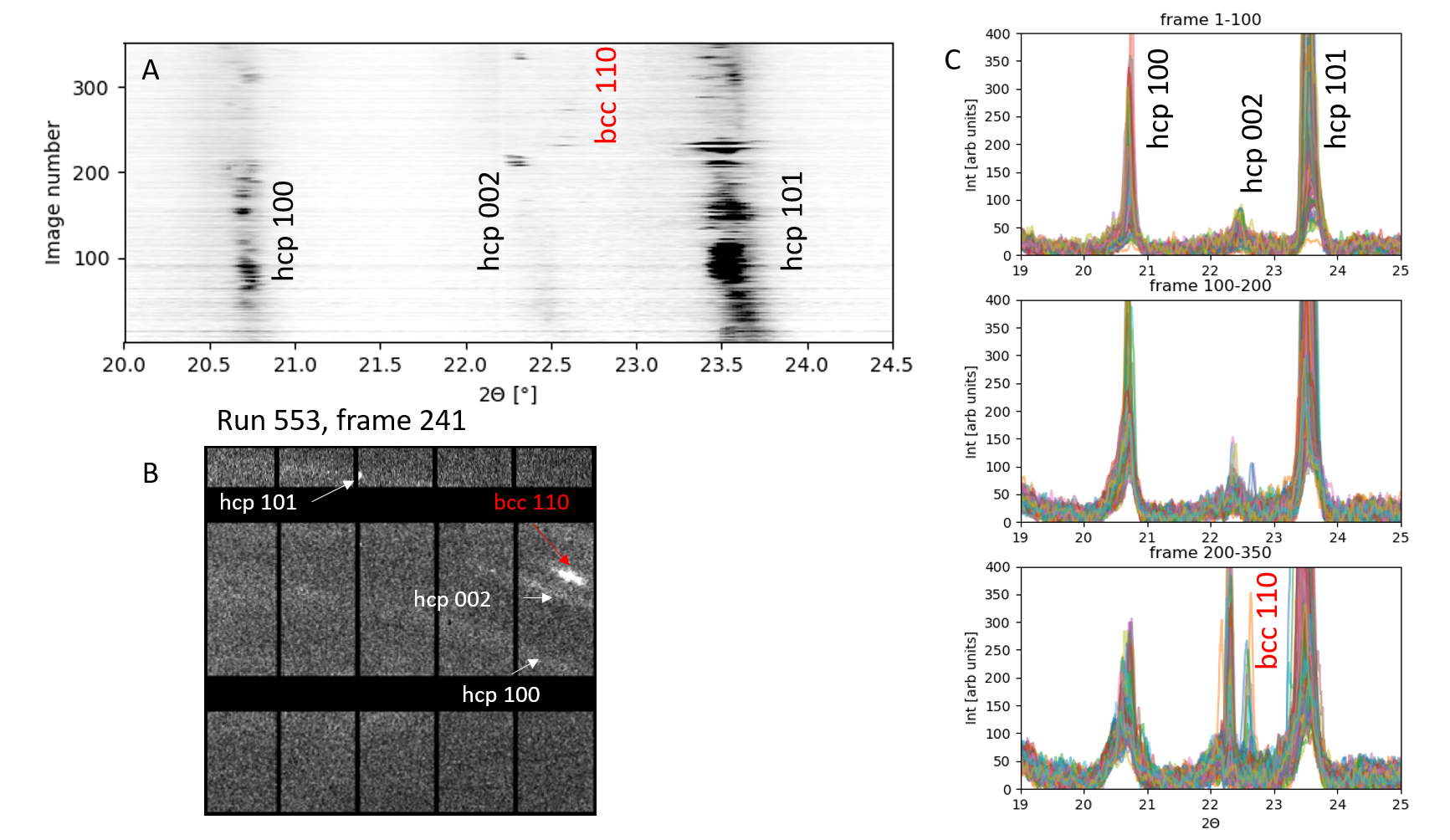}

\caption{\textbf{Data from the HIBEF38 DAC.} Run 553, at 8\% transmission, is the first train on this sample that transforms part of the sample into the cubic structure. \textbf{a}, Stacked integrated diffraction patterns from all 352 pulses showing a faint bcc reflection towards the end of the train (time from bottom-up), at the same time as disappearance of all hcp-Fe peaks. \textbf{b}, Focus on part of the raw image showing a BCC reflection. \textbf{c}, 2$\theta$ vs. intensity plots of the integrated patterns from the first 100 pulses (top), 100-200 pulses (center) and 200-350 pulses (bottom) showing the emerging 110 bcc-Fe reflection.}
\label{Fig 553}
\end{figure}

\begin{figure}%
\centering
\includegraphics[width=0.6\textwidth]{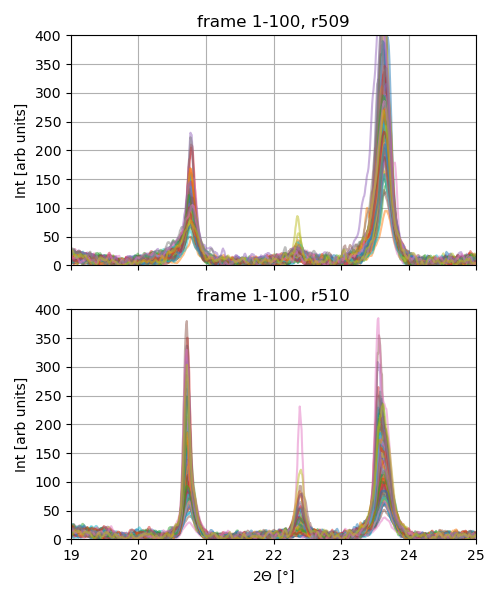}
\caption{\textbf{Bcc structure appears only at high temperatures.} The 110 bcc-Fe peak is not observed in the cold starting material before (first 100 pulses in r509; top) and after (first 100 pulses in r510, bottom) its appearance in the diffraction data. These successive heating runs were performed on the same sample position.}
\label{before_after_bcc}
\end{figure}

\subsection*{Temperature measurement, SOP}\label{sop}

Spectrally resolved streaked optical pyrometry (SOP) was used to monitor the temporal evolution of temperature during X-ray pulse trains by analyzing the spectral response of visible light emitted from the sample-X-ray interaction point \cite{Ball2023}. The SOP detection system comprises a visible spectrometer and a streak camera. The interaction point on the sample is imaged onto the entrance plane of the spectrometer, where the emitted light is dispersed by a grating within the 500-800 nm range onto the streak photocathode. A streak image is generated by sweeping the signal in time, allowing for the detection of variations in the intensity of light emitted over the duration of the X-ray train.

\begin{figure}%
\centering
\includegraphics[width=0.9\textwidth]{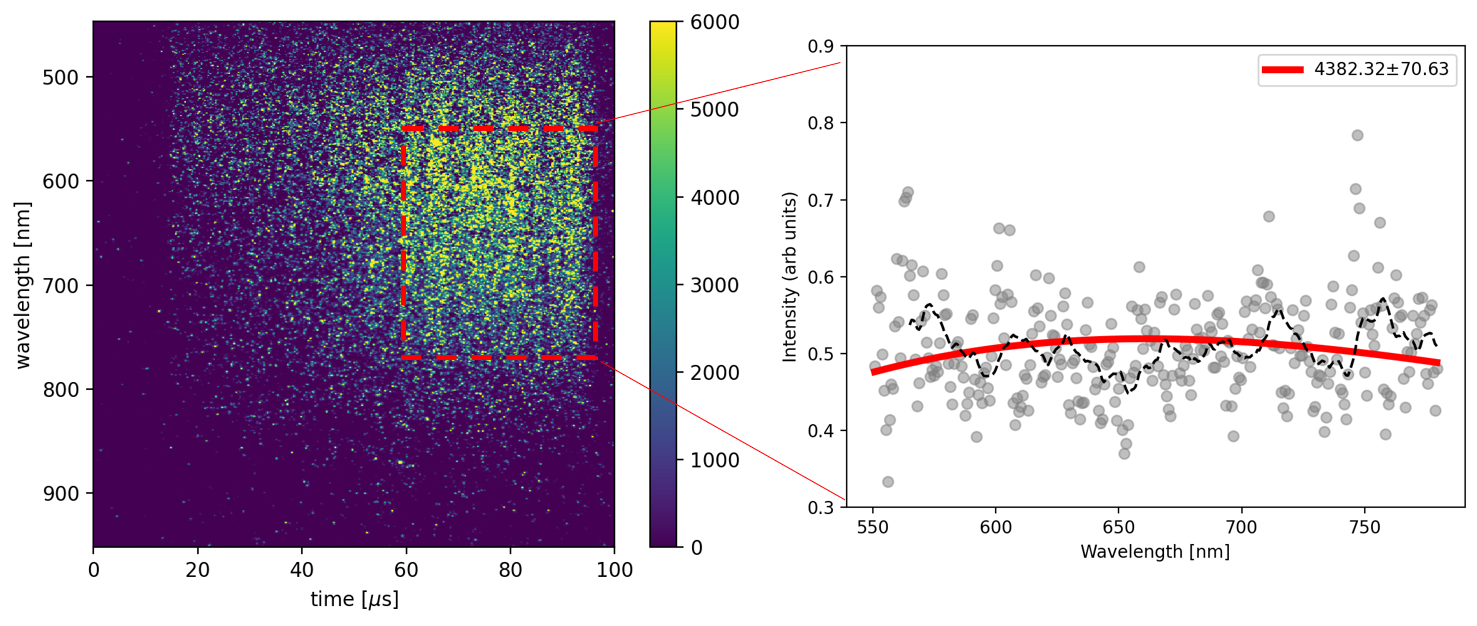}

\caption{\textbf{Spectrally resolved streak optical pyrometry (SOP) for HIBEF38, run 553.} A raw streak image (left) and temperature fit to emission data averaged over 60-95 $\mu$s time window (right).}\label{SOP_r553}
\end{figure}

In our experiments, we typically utilized 100 $\mu$s sweep windows to capture the entire pulse train consisting of 352 pulses at 4.5 MHz. Figure \ref{SOP_r553} displays the raw streak images from run 553, where the first BCC peaks were observed in HIBEF38. In addition to the thermal emission from the hotspot, fluorescence light from the sample, pressure-transmitting medium (PTM) and diamonds is also commonly observed. The fluorescence signal typically increases semi-linearly with the total X-ray power, whereas the thermal signal should rise much faster with power because of its proportionality to temperature\cite{Ball2023}. 
For HIBEF38, runs up to 551 are assumed to show primarily fluorescence emission from the diamond anvils, with an onset of thermal emission observed in runs 552 and 553. Run 551 then serves as the background image, which can be subtracted from run 553 to correct for fluorescence. An averaged signal from the second half of the train (time 60-95 $\mu$s, pixel 300-480) yields a temperature of 4382$\pm$70 K when fitting the Planck function within the wavelength range of 550-780 nm (Figure \ref{SOP_r553}). 

\subsection*{Finite element modeling}\label{FEM}

The structural information of high-pressure Fe at elevated temperatures is only accessible during cooling from previously induced higher temperatures by the preceding X-ray pulse. These temperatures immediately after X-ray energy absorption may significantly exceed the temperature detected at the time of probing. To visualize the dynamic temperature changes inside the sample during the pulse train, we employed finite element modeling using COMSOL Multiphysics, similar to the approach described by Meza-Galvez et al.\cite{MezaGalvez2020}. The model simulated heat conduction in a 2D axisymmetric cylindrical geometry, including a 2 $\mu$m thick Fe sample isolated from diamonds by 1 $\mu$m thick KCl layers. The X-ray spot was approximated by a Gaussian function with a sigma of 3 $\mu$m. The pulse energies of all 352 pulses were derived from measurements of scattered light from a thin diamond window by fast diodes placed before the interaction point, calibrated to absolute pulse energies using X-ray gas monitors (XGM)\cite{Maltezopoulos2019} prior to the experiment.
The material properties involved in the models are summarized in Table \ref{tab:comsol}. 

\begin{figure}
\centering
\includegraphics[width=0.6\textwidth]{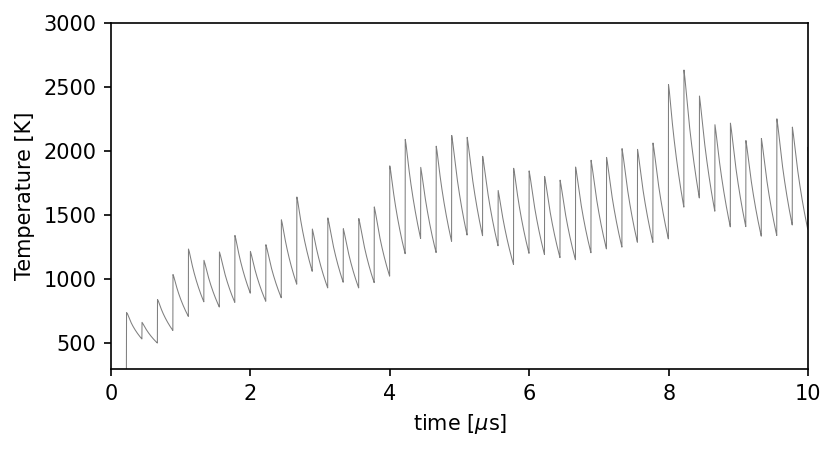}
\caption{Temporal temperature evolution at the center of the sample, zoomed into the first 10 microseconds, highlighting the oscillations caused by pulse absorption and conductive cooling.}
\label{T_zoom}
\end{figure}

\begin{figure}%
\centering
\includegraphics[width=0.6\textwidth]{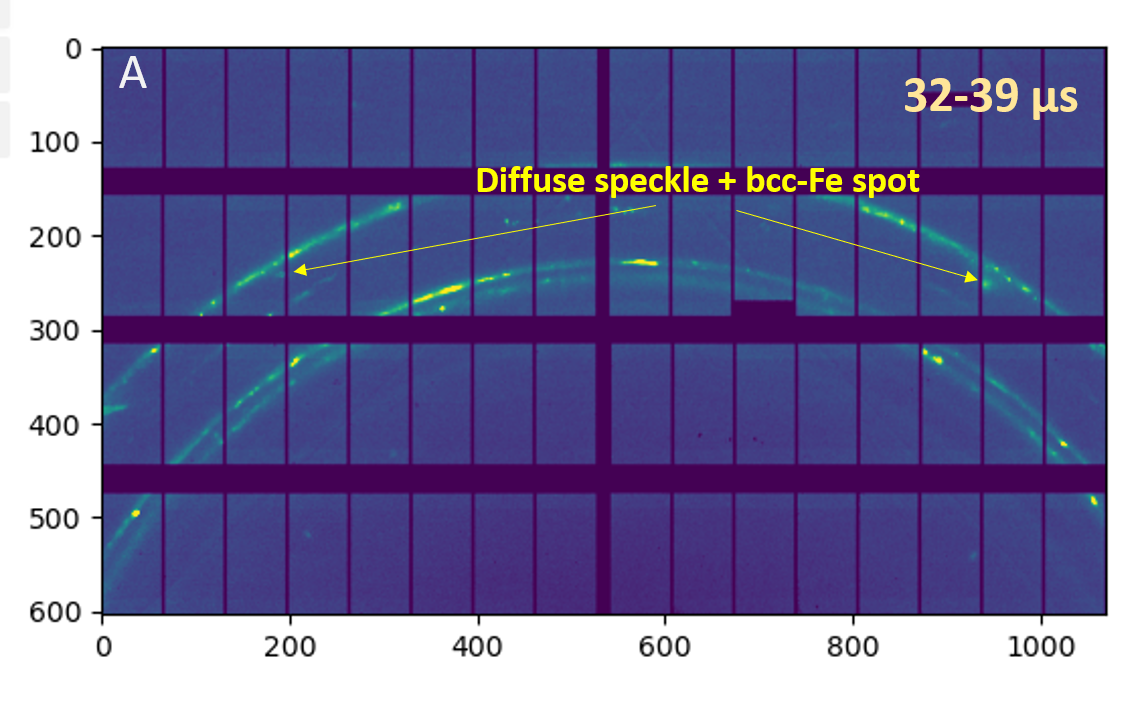}
\includegraphics[width=0.6\textwidth]{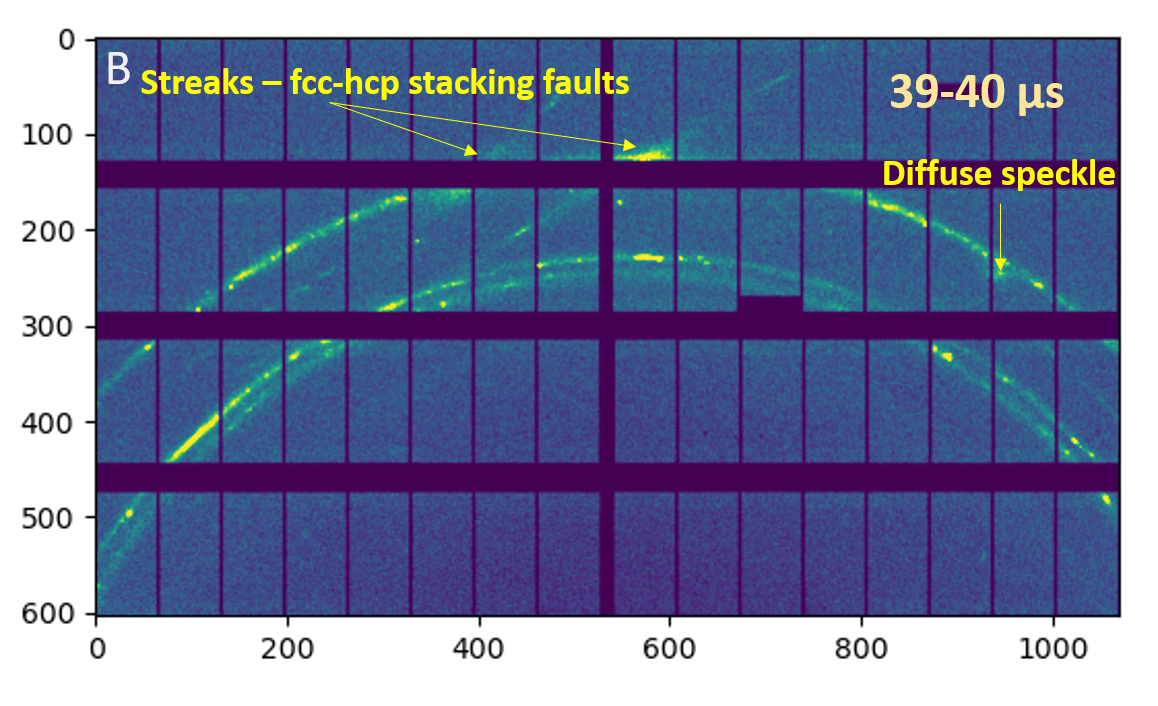}
\includegraphics[width=0.6\textwidth]{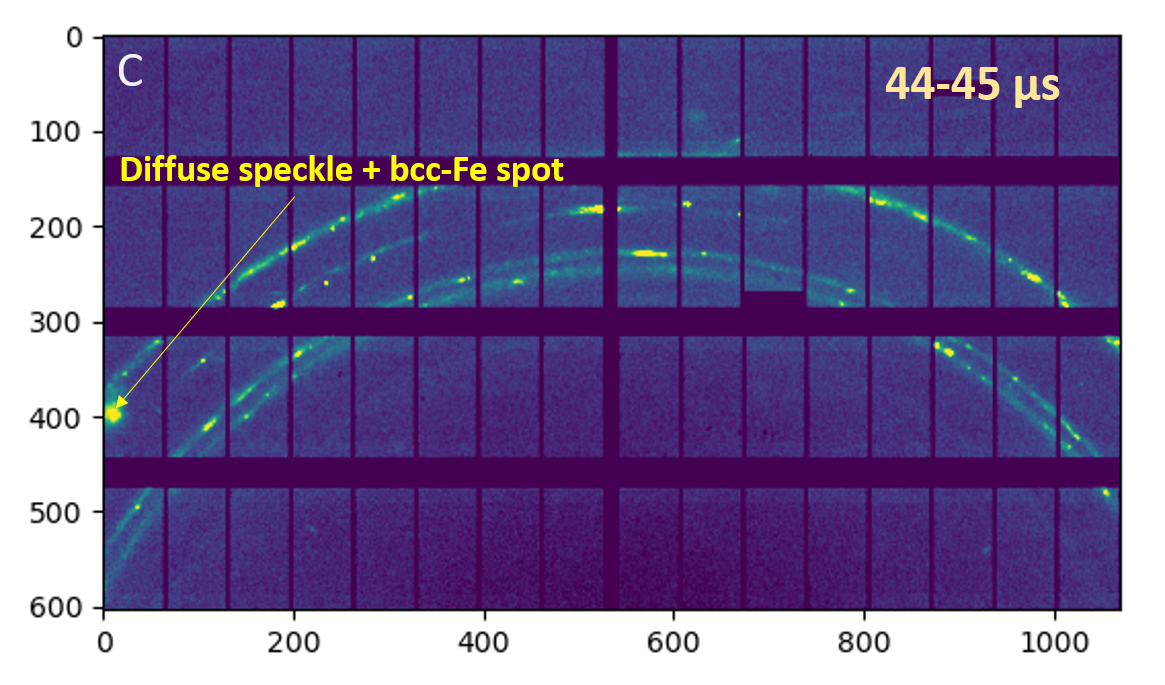}

\caption{
\textbf{Data from pressures below 180 GPa (BetsaA sample, 7\% X-ray transmission, run 452), showing transients states of bcc and fcc structures. }
\textbf{a}, Raw image averaged over frames 144-176 shows diffuse speckles with central Bragg spots. \textbf{b}, Averaged image over frames 176-181 with clear diffraction streaks due to fcc-hcp stacking faults. 
Diffuse speckle is here also visible on the right. \textbf{c}, Average image over frames 197-203 in the bottom panel shows an intense bcc-Fe reflection.}

\label{SF}
\end{figure}

The representative temperature distribution within the sample and the pressure transmitting medium (PTM) with such thin layer thicknesses is illustrated in Fig. \ref{Fig 509}d and \ref{T_zoom}. The temperature gradients between the center of the sample and the diamond interface are notably steep. The fluctuating temperature profile in the center (red) and surface (blue) of the Fe foil induced by the pulse train is also shown. Given an average temperature range of ~4400 K (as measured by SOP from the surface of the sample) during the final phase of the pulse train, it suggests that peak temperatures likely exceeded 6000 K, potentially leading to re-crystallization from a molten state.

\renewcommand{\thetable}{S\arabic{table}}
\begin{table}
    \caption{\label{tab:comsol}%
\textbf{Properties of materials included in the FEA simulation of the X-ray heated DAC.} $k$ - thermal conductivity; $\rho$ - density; $cp$ - specific heat capacity; $\mu$ - X-ray attenuation coefficient. }
\begin{ruledtabular}
    \begin{tabular}{ccccc}
         & Diamond & Re & Fe & KCl \\
        $k$ [W/m/K] & 990 & 48 & 50 & 20\\
        $\rho$ [kg/m$^3$]& 3515 & 21020 & 13000 & 5000\\
        $cp$ [J/kg/K] & 520 & 140 & 700 & 690\\
        $\mu$ [1/m] & 158  & --  & 43000  & 2513  \\
    \end{tabular}
  \end{ruledtabular}  
   
\end{table}

\begin{table}
\caption{\textbf{Pressure calculation.} Pressures were calculated from the thermal equation of state of Dewaele et al\cite{Dewaele2006} for hcp-Fe based on measured volumes from the HIBEF38 cell, run 553 (top) and CC273 cell, run 509 (bottom). Temperatures between 4000 K and 5000 K are considered.}
    \label{tab:EoS}
    \begin{ruledtabular}
    \begin{tabular}{cccc}
         V0 & V & T [K] & P [GPa] \\
         \hline
        14.6 &  & 300 & 209\\
             & 15.0 & 4000 & 221 \\
             & 15.0 & 5000 & 233 \\
        \hline
        14.7 &      & 300  & 202 \\
             &  14.8 & 4000 & 233 \\
             & 14.8 &  5000 & 246 \\
    \end{tabular}
   \end{ruledtabular} 
\end{table}

\clearpage 

\paragraph{Caption for Movie S1.}
\textbf{AGIPD XRD images and corresponding X-ray pulse energy.}
Movie of 352 2D diffraction images (left) created by X-ray pulses with energies depicted on the right. Data are from run 509, sample CC273, showing formation of bcc Fe at around 230 GPa and 4000-5000 K. 

\paragraph{Caption for Movie S2.}
\textbf{AGIPD XRD images and corresponding X-ray pulse energy.}
Movie of 352 2D diffraction images (left) created by X-ray pulses with energies depicted on the right. Data are from run 452, sample BetsaA, showing stacking faults and diffuse speckles at pressures above 236 GPa and high temperatures.

\bibliography{apssamp}

\end{document}